\documentclass[12pt]{article}
\usepackage{mathrsfs}
\usepackage{fullpage}
\usepackage{color}
\usepackage{amsfonts}
\usepackage{graphicx}
\usepackage{mathrsfs}
\usepackage{amsmath}
\usepackage{amssymb}
\usepackage{float}
\usepackage{rotating}
\allowdisplaybreaks
\usepackage[authoryear]{natbib}

\renewcommand\arraystretch{0.3}

\newtheorem{theorem}{Theorem}
\newtheorem{lemma}{Lemma}
\newtheorem{coro}{Corollary}

\def\proof {{\noindent\bf Proof.}\quad}
\def\var{\mathrm {var}}

\def\diag{\mathrm {diag}}

\newcommand{\bm}{\boldsymbol}

\def\tr{\mathrm {tr}}

\def\bu{\bm \upsilon}

\def\B{{\bf B}}
\def\z{{\bm z}}

\def\R{{\bf R}}
\def\I{{\bf I}}

\def\P{{\bf M}}

\def\X{{\bf F}}

\def\f{{\bf f}}
\def\W{{\bm W}}

\def\O{{\bf \Omega}}
\def\D{{\bf D}}

\def\tr{\mathrm {tr}}

\def\bms{{\bm\Sigma}}

\def\cd{\mathop{\rightarrow}\limits^{d}}
\def\mR{\mathcal{R}}

\def\bmv{\bm \varepsilon}

\newcommand{\be}{\begin{equation}}
\newcommand{\ee}{\end{equation}}
\newcommand{\beaa}{\begin{eqnarray*}}
\newcommand{\eeaa}{\end{eqnarray*}}
\newcommand{\bea}{\begin{eqnarray}}
\newcommand{\eea}{\end{eqnarray}}
\newcommand{\lbl}{\label}

\def\proof {{\noindent\bf Proof.}\quad}
\def\var{\mathrm {var}}

\def\diag{\mathrm {diag}}

\def\tr{\mathrm {tr}}

\def\B{{\bf B}}
\def\z{{\bm z}}

\def\R{{\bf R}}
\def\I{{\bf I}}

\def\P{{\bf M}}

\def\f{{\bf f}}
\def\W{{\bm W}}

\def\O{{\bf \Omega}}
\def\D{{\bf D}}

\def\tr{\mathrm {tr}}

\def\bms{{\bm\Sigma}}

\def\cd{\mathop{\rightarrow}\limits^{d}}
\def\mR{\mathbb{R}}

\def\bmv{\bm \varepsilon}

\title{Adaptive Strategy of Testing Alphas in High Dimensional Linear Factor Pricing Models}
\author{Chenxi Zhao, Ping Zhao, Long Feng and Zhaojun Wang\\
School of Statistics and Data Science, KLMDASR, LEBPS, and LPMC,\\ Nankai University}
\date{\today}
\def\boxit#1{\vbox{\hrule\hbox{\vrule\kern6pt  \vbox{\kern6pt#1\kern6pt}\kern6pt\vrule}\hrule}}

\begin{document}
\maketitle

\begin{abstract}
In recent years, there has been considerable research on testing alphas in high-dimensional linear factor pricing models. In our study, we introduce a novel max-type test procedure that performs well under sparse alternatives. Furthermore, we demonstrate that this new max-type test procedure is asymptotically independent from the sum-type test procedure proposed by Pesaran and Yamagata (2017). Building on this, we propose a Fisher combination test procedure that exhibits good performance for both dense and sparse alternatives.
\end{abstract}

\section{Introduction}
Linear Factor Pricing Model (LFPM) plays a central role in modern theories of security
pricing  \citep{Zhou1993Asset}. Let $N$ be the number of securities, and $T$ be
 the time dimension of the return series of each security. Let the
 $t$th return of the $i$th security be $Y_{it}$.  LFPM has
 the form
\begin{align}\label{eq:model}
Y_{it}=\alpha_i+\bm\beta_i^\top \f_t+\varepsilon_{it},
\end{align}
for $i=1,\dots, N$, $t=1, \dots, T$,
where
 $\f_t\equiv(f_{t1}, \cdots, f_{tp})^\top\in\mathcal{R}^p$ contains
 $p$ economic factors at time
 $t$,  $\alpha_i$ is a scalar representing the security specific
intercept, $\bm\beta_i\equiv(\beta_{i1},\cdots,\beta_{ip})^\top\in{\mathcal{R}}^p$
is a vector of multiple regression betas with respect
to the $p$ factors and {$\varepsilon_{it}$} is the corresponding
idiosyncratic error term with $Cov(\bmv_{t})\equiv\bms=(\sigma_{ij})_{N\times N}$, where
$\bmv_{t}\equiv(\varepsilon_{1t}, \cdots,
\varepsilon_{Nt})^\top\in\mR^N$ for each $t\in \{1,\cdots,T\}$. Many well known factor models belong to LPFM, such as the Sharpe-Lintner Capital Asset Pricing Model (CAPM) \citep{Sharpe1964,Lintner1965}, the Fama–French three-factor model \citep{Fama1993Common}, and the Fama–French five-factor model \citep{Fama2015A}.
The LFPM aims to clarify the variations in anticipated security returns by considering the interaction between security betas and systematic economic factors. It specifically predicts a linear relationship between the expected security return and the economic factors that are unique to each security. This linear framework is highly intuitive and offers the practical advantage of simplifying the modeling of security returns.

The intercept term $\alpha_i$ in (\ref{eq:model}) captures the
excessive return of the $i$th security. That is, other than the return
associated with the overall market factors, some securities may have
systematic positive or negative returns due to characteristics of the
individual securities, termed excess returns. Thus, the following pair of hypotheses
\begin{align} \label{h1}
H_0: \bm \alpha=\mathbf{0}~~ \text{versus}~~ H_1: \bm \alpha\not=\mathbf{0}
\end{align}
with $\bm \alpha\equiv (\alpha_1, \cdots, \alpha_N)^\top=\bm 0$
allows to assess whether the excess return of the market portfolio presents. When the number of securities $N$ is fixed, many test procedures are devised under the normal or non-normal distribution assumptions, such as \citet{GIBBONS1989A}, \citet{Mackinlay1991Using}, \citet{Zhou1993Asset} and \citet{Beaulieu2007Multivariate}, etc. Nowadays, thousands of securities are traded in modern financial markets. So the assumption of fixed $N$ is not appropriate.

Many efforts have been devoted for testing alphas in high dimensional linear factor pricing model. Generally speaking, there are three-types test procedures in literature. The first type is the sum-type test procedure, which are constructed by summing the square of the classic t-test statistic of each security, such as \citet{Pesaran2017Testing}, \citet{GagliardiniTime},\citet{ma2020testing} and \citet{giglio2021thousands}, etc. These sum-type tests have good performance against dense alternatives, i.e. the alphas of over a large number of securities are not zero. The second type is the max-type test procedure, which are constructed by taking the maximum of the squared standard t-ratio of each securities, such as \citet{feng2022high}. The max-type test procedure is powerful under the sparse alternative, i.e. only a few alphas of securities are not zero. However, we can not know whether the alternative is dense or sparse in real applications. So the third type of test procedure are constructed by combining the above sum-type and max-type test procedures together. \cite{Fan2015Power} proposed a power enhancement procedure by adding sum-type test statistics and max-type test statistics. \citet{feng2022high}, \citet{yu2019innovated} proposed a Fisher combination test based on the asymptotically independence between the sum-type test statistic and the max-type test statistic. In this paper, we will propose a novel adaptive strategy test procedure, which belongs to the third type test procedure.

\cite{yu2019innovated} employed the thresholding covariance matrix estimator of \citet{Fan2015Power} and proposed a novel sum-type test statistic in their Fisher combination test procedure. However, there would be a non-negligible bias term in their sum-type test statistic if the covariance matrix estimator is not very accurate. In addition, the max-type test statistic in \citet{feng2022high} and \cite{yu2019innovated} do not consider the information of the correlation of each security, which may perform not very well under some special sparse alternatives. To overcome this issue, we first proposed a novel max-type test statistic by standardizing the vector of t-test statistics of each security. Then, we prove that the new max-type test statistic is asymptotically independent with the sum-type test statistic of \citet{Pesaran2017Testing}. Finnally, we propose a new Fisher combination test by the above two asymptotically independent test statistics. Simulation studies show the superior of our proposed procedure.

The rest of paper is organized as follows. In Section 2, we introduce the new max-type test statistic and establish the theoretical results. And we also propose an adaptive strategy testing procedure. Simulation studies are presented in Section 3. All the technical proofs are given in the Appendix.

\section{Our test procedure}

\subsection{Max-type test}
To test the hypothesis \eqref{h1}, \cite{feng2022high} propose a max-of-squares type test, named the MAX test, with the
test statistic constructed as
\begin{align}\label{max1}
M_{\I}=\max_{1\le i \le N} t_i^2,
\end{align}
where
\begin{align*}
t_i=\frac{\hat{\alpha}_i\sqrt{(\bm 1_T^\top \P_{\X}\bm
  1_T)}}{\sqrt{v^{-1}\hat{\bmv}_{i\cdot}^\top\hat{\bmv}_{i\cdot}}}
\end{align*}
 is the squared standard t-ratio of $\alpha_i$ in the OLS regression
 of
$y_{it}$ on an intercept and $\f_t$,  and $v=T-p-1$.
Here, $\bm 1_T=(1,\cdots,1)^\top$, $\I_T$ is the $T\times T$ identity matrix, $\X=(\f_1, \cdots, \f_T)^\top\in\mR^{T\times p}$ and $\P_{\X}=\I_T-\X(\X^\top\X)^{-1}\X^\top$;
$\hat{\bm \alpha}=(\hat{\alpha}_1,\cdots,\hat{\alpha}_N)^\top$ is the
OLS estimator of $\bm \alpha$, where
$\hat{\alpha}_i=\mathbf{Y}_{i\cdot}^\top \P_{\X}\bm 1_T/(\bm 1_T^\top
\P_{\X}\bm 1_T)$
and ${\bf Y}_{i\cdot}=(Y_{i1}, \cdots, Y_{iT})^\top\in\mR^T$;
$\hat{\bmv}_{it}$ is the OLS residual from the regression of $y_{it}$
on an intercept and $\f_t$,
$\bmv_{i\cdot}\equiv(\varepsilon_{i1}, \cdots, \varepsilon_{iT})^\top\in\mR^T$ and $\hat{\bmv}_{i\cdot}\equiv(\hat{\varepsilon}_{i1}, \cdots, \hat{\varepsilon}_{iT})^\top=\P_{\X}(\mathbf{Y}_i-\hat{\alpha}_i)$.

%
%

However, the test statistic (\ref{max1}) do not consider the correlation between those securities. Under the null hypothesis, $\bm \alpha=\bm 0$ is equivalent to ${\bf B}\bm \alpha=\bm 0$ for any positive definite matrix ${\bf B}$. Under the null hypothesis, we have $\bm t=(t_1,\cdots, t_N)\cd N(\bm 0, \R)$ if $\bmv_t\sim N(0,\bms)$ where $\bms$ is the covariance matrix and $\R$ is the corresponding correlation matrix. So a nature choice of ${\bf B}$ is $\R^{-1/2}$ which standardized the $t$-test statistics $\bm t$. So, if we have a good consistent estimator of $\O^{1/2}\doteq \R^{-1/2}$, $\hat{\O}^{1/2}$, we could propose a new max-type test
\begin{align}\label{max2}
M_{\hat{\O}^{1/2}}=\max_{1\le i \le N} \nu_i^2,
\end{align}
where $\nu=(\nu_1,\cdots,\nu_N)^\top=\hat{\O}^{1/2}\bm t$.

Next, we consider the theoretical properties of $M_{\hat{\O}^{1/2}}$ based on the following assumptions.
\begin{itemize}
\item[(A1)] $\boldsymbol{\varepsilon}_{1}, \ldots, \boldsymbol{\varepsilon}_{T}$ are independently and identically distributed with $\bmv_t=\bms^{1/2}\bm\xi_t$. And we assume $\bm \xi_t=(\xi_{1t},\cdots,\xi_{Nt})$ contains independent components $\xi_{it}$'s with $E(\xi_{it})=0$ and $\var(\xi_{it})=1$, $E(\xi_{it}^4)<c$ for some positive constant $c$. Define $\R=\D^{-1/2}\bms\D^{-1/2}=(r_{ij})_{1\le i,j\le N}$ where $\D$ is the diagonal matrix of $\bms$. (i) There exists $c_{3}>0$, such that $c_{3}^{-1} \leq \lambda_{\min }(\mathbf{R}) \leq \lambda_{\max }(\mathbf{R}) \leq c_{3} .$ (ii) There exists $r_{1}>0$, such that $\max_{1 \leq i<j \leq N}\left|r_{i j}\right| \leq$ $r_{1}<1$.
\item[(A2)]     $\xi_{it}$'s have one of the following three types of tails: (i) sub-Gaussian-type tails, i.e.
there exist $\eta>0$ and $K>0$ such that
  $E\{\exp(\eta \xi_{it}^2/\sigma_{ii})\}\le K$ for $1\le i\le N$,
 where $N$ satisfies
  $\log(N)=o(T^{1/4})$; (ii) sub-exponential-type tails, i.e.
there exist $\eta>0$ and $K>0$ such that
  $E\{\exp(\eta |\xi_{it}|/\sigma_{ii}^{1/2})\}\le K$ for $1\le i\le N$,
 where $N$ satisfies
  $\log(N)=o(T^{1/4})$; (iii)  sub-polynomial-type tails, i.e.
for some constants $\gamma_0$,
  $\epsilon>0$ and $K>0$,
  $E|\xi_{it}/\sigma_{ii}^{1/2}|^{2\gamma_0+2+\epsilon}\le K$ for $1\le i\le N$,
where $N$ satisfies $N\le c_1 T^{\gamma_0}$
for some constants
  $c_1> 0$.
\item[(A3)] (i) $\left\{\mathbf{f}_{1}, \ldots, \mathbf{f}_{T}\right\}$ is strictly stationary and independent of $\left\{\boldsymbol{\varepsilon}_{1}, \ldots, \boldsymbol{\varepsilon}_{T}\right\}$. (ii) $\mathbf{f}_{t}^{\prime} \mathbf{f}_{t} \leq K<\infty$, for all $t .$ The $(m+1) \times(m+1) \operatorname{matrix} T^{-1} \mathbf{G}^{\prime} \mathbf{G}$, with $\mathbf{G}=\left(\boldsymbol{1}_{T}, \mathbf{F}\right)$, is a positive definite matrix for all $T$, and as $T \rightarrow \infty$, and $\bm 1_{T}^{\prime} \mathbf{M}_{F} \bm 1_{T}>0$ where $\mathbf{M}_{F}=\mathbf{I}_{T}-\mathbf{F}\left(\mathbf{F}^{\prime} \mathbf{F}\right)^{-1} \mathbf{F}^{\prime} .$ (iii) $\operatorname{cov}\left(\mathbf{f}_{t}\right)$ is strictly positive definite.
\item[(A4)] We assume that the estimator $\hat{\O}^{1/2}=(\hat{\omega}_{ij})$ has at least a logarithmic rate of convergence
\begin{align*}
\|\hat{\Omega}^{1/2}-\Omega^{1/2}\|_{L_{1}}=o_{{p}}\left\{\frac{1}{\log (N)}\right\},\max _{1 \leqslant i \leqslant p}\left|\hat{\omega}_{i, i}-\omega_{i, i}\right|=o_{p}\left\{\frac{1}{\log (N)}\right\}
\end{align*}
\end{itemize}

Assumption (A1) assume that the error term follows the independent component model, which is widely used in high dimensional data analysis, such as \cite{li2019testing,chen2022asymptotic}. The assumption of the correlation matrix is the same as condition 1 and 2 in \cite{Cai2014}. Assumption (A2) consider three types of tails of $\xi_{it}$, which allows the dimension $N$ smaller as the tails gets heavier. Assumption (A3) is also widely assumed in high dimensional linear factor pricing model, such as \cite{Fan2015Power,feng2022high}. Assumption (A4) is the same as condition (8) in \cite{Cai2014}, which is a rather weak requirement on $\hat{\O}$ and can be easily satisfied by many high dimensional precision matrix estimators, such as \cite{bickel2008covariance,cai2011constrained}.

\begin{theorem}\label{tho1}
Under the condition (A1)-(A4),  we have
\begin{align}
P_{H_0}\left(M_{\hat{\O}^{1/2}}-2\log(N)+\log\log (N)\le x\right)\to F(x)\doteq \exp\{-(1/\sqrt{\pi})\exp(-x/2)\}
\end{align}
for any $x\in \mathbb{R}$.
\end{theorem}

Based on Theorem \ref{tho1}, a level $\gamma$ test will then be performed through
rejecting $H_0$ when $M_{\hat{\O}^{1/2}}-2\log(N)+\log\{\log(N)\}$ is larger
than the $1-\gamma$
quantile, $q_{\gamma}=-\log(\pi)-2\log\{\log(1-\gamma)^{-1}\}$, of the type I extreme value distribution.

Next, we will show the consistency of the new proposed max-type test procedure.
Define
\begin{align*}
	\mathcal{S}(k_N) \equiv \{\bm{\alpha} \in \mathcal{R}^N: \sum_{I=1}^N I(\alpha_i \neq 0)=k_N\}
\end{align*}

Here, $\mathcal{S}(k_N) $ denote the set of $k_N$-sparse vectors with $k_N=O(N^{r})$ and $r<\frac{1}{4}$.
\begin{theorem}\label{tho2}
	Under the condition (A1)-(A4),  as $min(T,N) \to \infty$
	\begin{align}
		\inf_{\alpha\in\mathcal{A}(\beta) \cap \mathcal{S}(k_N)} P(\phi_{\gamma}=1) \to 1
	\end{align}
where $\phi_\gamma \equiv I[M_{{\hat{\O}}^{1/2}}-2\log(N)+\log\log (N)\ge q_{\gamma}]$, $\beta \ge 1/\min_i\{\omega_{i, i}^2\}+\epsilon$ for some constant $\epsilon>0$, and
\begin{align*}
	\mathcal{A}(\beta) \equiv \{ \bm{\alpha} \equiv (\alpha_1,\dots,\alpha_N)\in \mathcal{R}^{N}:\max_{1 \leq i<j \leq N} \vert \alpha_i/\sigma_{ii}^{1/2}\vert \ge \sqrt{8\beta\log(N)/T}\}
\end{align*}
\end{theorem}

\subsection{Adaptive Test}
As shown in Theorem \ref{tho2}, the max-type test statistic $M_{\hat{\O}^{1/2}}$  has good performance under sparse alternatives. While for the dense alternatives, \cite{Pesaran2017Testing} proposed a sum-type test statistic:
\begin{align*}
T_{{\rm PY}}=\frac{N^{-1/2}\sum_{i=1}^N
  \{t_i^2-v/(v-2)\}}{v/(v-2)\sqrt{2(v-1)/(v-4)\{1+(N-1)\tilde{\rho}_{MT}^2\}}},
\end{align*}
where $v=T-p-1$ and
$\tilde{\rho}_{MT}=2/\{N(N-1)\}\sum_{i=2}^N\sum_{j=1}^{i-1}\tilde{\rho}^2_{ij}$
is the corresponding correlation estimator of
$\rho^2=2/\{N(N-1)\}\sum_{i=2}^N\sum_{j=1}^{i-1}\rho^2_{ij}$
with $\tilde{\rho}_{ij}$ denoting the multiple testing estimator of
the correlation
$\rho_{ij}=\sigma_{ij}/(\sigma_{ii}^{1/2}\sigma_{jj}^{1/2})$
\citep{bailey2019a}.
%
Under the null hypothesis, we have $T_{PY}\cd N(0,1)$. Thus, we will reject the null hypothesis when $T_{PY}>z_{\gamma}$ where $z_{\gamma}$ is the upper $\gamma$ quantile of the standard normal distribution.

In practice, we can not know whether the alternative is dense or sparse. So we proposed the following Fisher combination test statistic
\begin{align}
T_{FC2}=-2\log p_{sum}-2\log p_{max2}
\end{align}
where
\begin{align}
p_{sum}=1-\Phi(T_{PY}), ~~ p_{max2}=1-F(M_{\hat{\O}^{1/2}}-2\log N+\log \log (N))
\end{align}

To establish the asymptotic distribution of $T_{FC2}$, we need the following additional conditions:
\begin{itemize}
\item[(A5)]  $\sup_{i,t}E(|\xi_{it}|^{8+c})\le K<\infty$ for some $c>0$.
\item[(A6)] Let $||\R^{1/2}||_1<K$ and $||\R^{1/2}||_{\infty}<K$.
\end{itemize}
Assumption (A5) and (A6) are the same as Assumption 3 in \cite{Pesaran2017Testing}. As shown in Remark 3 in \cite{Pesaran2017Testing}, the sparse conditions of $\R$ is particularly important in finance where security returns could be affected by weak unobserved factors. Based on these assumptions, we can establish the asymptotic independence between the sum-type test statistic $T_{PY}$ and the max-type test statistic $M_{\hat{\O}^{1/2}}$.

\begin{theorem}\label{tho3}
Under the conditions (A1), (A3)-(A6), we have, if $N/T^2\to 0$,
\begin{align}\label{asy2}
P\left(T_{PY}<x,M_{\hat{\O}^{1/2}}-2\log N+\log\log N<y\right)\to \Phi(x)F(y)
\end{align}
\end{theorem}
Based on Theorem \ref{tho3}, we have the limit distribution of $T_{FC2}$ as follow.
\begin{coro}
Under the same condition as Theorem \ref{tho3}, we have $T_{FC2}\cd \chi_4^2$ under the null hypothesis.
\end{coro}

Because the convergence rate of $T_{FC2}$ is not very fast, we suggest to use the critical value $(1+\log^{-1}(TN^{1/2}))\chi_4^2(\gamma)$ in practice, where $\chi_4^2(\gamma)$ is the upper $\gamma$-quantile of $\chi_4^2$ distribution.

We consider the following alternative hypothesis:
\begin{align}\label{h12}
H_1: \tilde\alpha_i\not=0, i\in \mathcal{M}, ~~ |\mathcal{M}|=m, ~~m=o(N^{1/2}), ~~N^{-1}T\tilde{\bm \alpha}^\top \tilde{\bm \alpha}=o(1)
\end{align}
where $\tilde{\bm \alpha}=\O^{1/2}\bm \alpha$. Next, we also demonstrate that $T_{PY}$ is still asymptotically independent with $M_{\hat{\O}^{1/2}}$ under this special alternatives.

\begin{theorem}\label{tho33}
Under the conditions (A1), (A3)-(A6), we have, if $N/T^2\to 0$,
\begin{align}\label{asy22}
P\left(T_{PY}<x,M_{\hat{\O}^{1/2}}-2\log N+\log\log N<y\right)\to P\left(T_{PY}<x\right)P\left(M_{\hat{\O}^{1/2}}-2\log N+\log\log N<y\right)
\end{align}
under the alternative hypothesis (\ref{h12}).
\end{theorem}
Similar to the arguments in \cite{wang2023}, we have
\begin{align*}
\beta_{T_{FC2}}\ge\beta_{sum, \gamma/2}+\beta_{max2,\gamma/2}-\beta_{sum, \gamma/2}\beta_{max2,\gamma/2}
\end{align*}
where $\beta_{sum}, \beta_{max2, \gamma}$ is the power function of the sum-type test $T_{PY}$ and $M_{\hat{\O}^{1/2}}$ at significant value $\gamma$, respectively. For a small $\gamma$, the difference between $\beta_{max2,\gamma}$ and $\beta_{max2,\gamma/2}$ should be
small, and the same fact applies to $\beta_{sum,\gamma}$. Consequently, the power of the adaptive test $T_{FC2}$ would be no smaller than or even significantly larger than that of either max-type or sum-type test.

\section{Simulation}

This experiment is designed to mimic the commonly used Fama-French
three-factor model,
where the factors $\f_t$ have strong serial correlation and heterogeneous variance.
Specifically, we consider a modified version of the example studied
in Section 5.1 of \citet{Pesaran2017Testing}.
The response $Y_{it}$ are generated according to the LFPM in
(\ref{eq:model}) with $p=3$:
\begin{align*}
Y_{it}=\alpha_i+\sum_{k=1}^p \beta_{ik} f_{kt} +\varepsilon_{it},
\end{align*}
where the three factors, $f_{t1}$, $f_{t2}$ and $f_{t3}$, are the
Fama-French three factors.
We generate each factor from an autoregressive conditional
heteroskedasticity process and the  GARCH(1,1) model respectively.
Specifically
The data generating process is as follows:
\begin{align*}
Y_{it}=\alpha_i+\sum_{k=1}^p \beta_{ik} f_{kt}
  +\varepsilon_{it},~i\in\{1,\cdots,N\},~t\in\{1,\cdots,T\}.
\end{align*}
Let $p=3$. The three factors $f_{1t}$, $f_{2t}$ and $f_{3t}$ are presented as
the Fama--French three factors, Market factor, SMB and HML.
We generate the factors as follows,  where for each factor,
the error term follows a GARCH$(1,1)$ process, and all the coefficients
are the same as that in \cite{Pesaran2017Testing}. Specifically,
\begin{align*}
f_{1t}=&0.53+0.06 f_{1,t-1}+ h_{1t}^{1/2}\zeta_{1t},~\mbox{Market factor},\\
f_{2t}=&0.19+0.19 f_{2,t-1}+ h_{2t}^{1/2}\zeta_{2t},~\mbox{SMB factor},\\
f_{3t}=&0.19+0.05 f_{3,t-1}+ h_{3t}^{1/2}\zeta_{3t},~\mbox{HML factor},
\end{align*}
where $\zeta_{kt}$'s are simulated from a standard normal
distribution, the variance terms $h_{kt}$'s are generated from
\begin{align*}
h_{1t}&=0.89+0.85 h_{1,t-1}+ 0.11 \zeta_{1,t-1}^2, ~\mbox{Market factor},\\
h_{2t}&=0.62+0.74 h_{2,t-1}+ 0.19\zeta_{2,t-1}^2, ~\mbox{SMB},\\
h_{3t}&=0.80+0.76 h_{3,t-1}+ 0.15 \zeta_{3,t-1}^2, ~\mbox{HML}.
\end{align*}
Similar to \cite{Pesaran2017Testing}, the above process is simulated
over the periods $t=-49, \cdots, 0, 1, \cdots, T$ with the initial
values
$f_{k,-50}=0$ and $h_{k,-50}=1$ for any $k=1,2$ and 3. We use the
simulated data for observations $t=1, \cdots, T$ for our final
experiments.

The errors are generated from $\bmv_{t}\sim \bms^{1/2} \z_{t}$, where
$\z_t=(z_{1t},\cdots,z_{Nt})^\top$. We consider four settings of
$z_{it}$'s:
\begin{itemize}
\item[(i)] Normal distribution: $z_{it}\overset{i.i.d}{\sim}
N(0,1)$;
\item[(ii)] $t_3$ distribution: $z_{it}\overset{i.i.d}{\sim}
t(5)/\sqrt{5/3}$
\item[(iii)] mixture normal distribution: $z_{it}\overset{i.i.d}{\sim}
\{0.9N(0,1)+0.1N(0,9)\}/\sqrt{1.8}$
\end{itemize}

\begin{table}[!hb]
\begin{center}
\caption{\label{t1} Sizes of tests with $T=100$.}
                     \vspace{0.5cm} \setlength{\tabcolsep}{3pt}
                     \renewcommand{\arraystretch}{1} {
\begin{tabular}{c|cccc|cccc|cccc|cccc}
\hline \hline
  &  \multicolumn{4}{|c}{Model 1} & \multicolumn{4}{|c}{Model 2}& \multicolumn{4}{|c}{Model 3}& \multicolumn{4}{|c}{Model 4}\\ \hline
  &  \multicolumn{4}{c}{$N$} & \multicolumn{4}{|c}{$N$} & \multicolumn{4}{|c}{$N$} & \multicolumn{4}{|c}{$N$}\\
 Method &50 & 100 & 200 & 300 & 50 & 100 & 200 & 300 & 50 & 100 & 200 & 300 & 50 & 100 & 200 & 300\\
\hline
\hline
\multicolumn{17}{c}{Normal Errors}\\\hline
PY&6.2&6.1&7.9&7.8&5.7&6.1&7.3&5.5&5.8&5.8&5.4&4.4&5.3&6.4&4.9&4.5\\
MAX1&3.1&4.6&4.6&4.6&4.1&3.5&4.2&5&3.2&5.3&5.5&5.2&3.6&5.3&4.5&4.9\\
MAX2&8.2&6.9&4.8&5.3&3.9&3.6&5&7.4&4.1&5.2&5.4&5.2&4&5.6&4.4&5\\
FC1&4.9&5.6&6.1&6.7&5.2&5.3&5.1&4&4.8&5.7&5&4.5&4.7&6.1&4.8&4.5\\
FC2&6.4&6.4&6.6&6.2&5.2&5.4&5.4&5.9&5.4&5.6&4.9&4.5&4.9&6.4&4.9&4.5\\
COM&5.1&4.9&5.6&6.8&5.1&5.2&6.5&5.1&4.9&5.7&5.3&4.2&3.9&6.1&5&4.1\\
PE&22.8&20.1&14.3&12.2&6.5&8.4&10.4&12&9.9&11.9&12.1&12.4&9&12.5&11&13.1\\
\multicolumn{17}{c}{$t(5)$ Errors}\\\hline
PY&7.8&6.9&6.8&5.7&5.2&4.1&4.6&4&5.7&4.4&6&5.7&6.2&5.3&5.5&4.5\\
MAX1&3.9&2.6&3.8&4.4&3&4.7&3.8&4.1&3.4&3.3&3.5&4.7&3.7&4.2&3.9&4.3\\
MAX2&8.1&5.4&5.2&4.4&2.9&4.6&4.7&6.5&3.5&3.3&3.6&4.9&3.7&4.3&4&4.4\\
FC1&6.8&5.2&6.6&5.4&3.7&3.4&3.5&3.4&4.7&3.7&4.6&5.5&5.6&4.4&3.9&3.4\\
FC2&8.6&5.8&6.5&4.8&3.7&3.4&3.9&4.9&4.9&3.5&4.6&5.5&5.3&4.2&4&3.6\\
COM&5.8&5.8&6.2&5.5&3.7&3.5&4.3&3.6&4.7&3.7&4.2&6.4&5.7&3.9&5.4&3.2\\
PE&23.1&19.3&16.9&9.6&5.7&5.9&9.6&11.6&9.3&9.5&11.9&12.2&10.1&10.4&10.9&10.9\\
\multicolumn{17}{c}{Mixture Normal Errors}\\\hline
PY&7.1&5.5&6.4&7.5&5.6&6.7&5.4&4.2&5.3&7&3.8&5.9&5.4&4.8&5.3&6.1\\
MAX1&2.7&3.4&4.5&4.1&2.3&4.3&2.2&4.3&3.3&4&4.7&3.3&3.6&3.8&4&3.3\\
MAX2&8.4&5.5&3.5&4.8&2.3&4.3&2.8&6.8&3.2&3.5&4.9&3.6&4.2&4.1&4.1&3.6\\
FC1&5.5&4.4&5.3&6&3.7&6.1&3.4&3.8&4.3&5.5&5&4.4&4.3&3.7&4.7&4\\
FC2&8.1&4.4&5.4&6.2&3.6&6.2&3.6&5.2&4.2&5.6&5.2&4.3&4.1&4&4.5&4.3\\
COM&5.5&4.6&5.2&6&3.7&5.8&4.3&3.7&4&5.6&4.8&5.3&4.2&3.7&4.7&4.2\\
PE&24.2&17.1&12.5&10.2&7.2&9.2&9.2&12.8&9.7&12.6&10&13.5&9.3&9.2&10.4&11.2\\
\hline
\hline
\end{tabular}}
\end{center}
\end{table}

We consider four models for the covariance matrix
\begin{itemize}
	\item[(1)] Model 1: $\bms=(0.7^{|i-j|})_{1\le i,j\le N}$;
	\item[(2)] Model 2: $\bms=\D^{1/2}\R\D^{1/2}$,
	$\D=\diag\{\sigma_1^2,\cdots,\sigma_N^2\}$ and $\R=\I_N+\bm b\bm b
	^\top-\tilde{\B}$, where $\bm b=(b_1,\cdots,b_N)^\top$ and
	$\tilde{\B}=\diag\{b_1^2,\cdots,b_N^2\}$.
	We generate $\sigma_{ii}\sim U(1,2)$. We randomly generate $[N^{0.3}]$ elements of $\bm b$ from $U(0.7,0.9)$ and set
	the remaining elements to be zero. Here $[\cdot]$
	denotes the integer part of a real number.
	\item[(3)] Model 3: $\bms=\O^{-1}$ where $\O=(0.6^{|i-j|})_{1\le i,j\le N}$;
	\item[(4)] Model 4: $\bms=\bm \gamma \bm
	\gamma^{T}+(\I_p-\rho_{\epsilon}\W)^{-1}(\I_p-\rho_{\epsilon}\W^{T})^{-1},$ where
	$\bm \gamma=(\gamma_1,\cdots,\gamma_{[p^{\delta_{\gamma}}]},0,0,\cdots,0)^{T}$.
	Here  $\gamma_i$ with $i=1,\cdots, [p^{\delta_{\gamma}}]$ are generated independently from $Uniform(0.7,0.9)$. Let
	$\rho_{\epsilon}=0.5$ and $\delta_{\gamma}=0.3$. Let $\W=(w_{i_1i_2})_{1\le i_1,i_2\le p}$ have a so-called rook form, i.e.,
	all elements of $\W$ are zero except that
	$w_{i_1+1,i_1}=w_{i_2-1,i_2}=0.5$ for $i_1=1,\cdots,p-2$ and
	$i_2=3,\cdots,p$, and $w_{1,2}=w_{p,p-1}=1$.
\end{itemize}

Here $PY$ means $T_{PY}$, MAX1 means $M_{\I}$, MAX2 means $M_{\hat{\O}^{1/2}}$, $HC1$ means $T_{FC1}$ which is the Fisher combination test based on $T_{PY}$ and $M_{\I}$, HC2 means $T_{FC2}$, COM means the combination method proposed by \citet{feng2022high}. PE means the power enhancement method proposed by \citet{Fan2015Power}. We estimate the covariance matrix by the thresholding method proposed by \citet{bickel2008covariance}. For fair comparison, we use the same covariance estimator in PE. Finally, the three groups of coefficients corresponding to the three factors, $\beta_{i1}$, $\beta_{i2}$ and $\beta_{i3}$,
are generated independently from $U(0.2,2)$, $U(-1,1.5)$ and $U(-1.5,1.5)$, respectively. We set $\bm\alpha=\bm 0$ under the null hypothesis. Table \ref{t1} reports the empirical sizes of each test with sample size $T=100$. We found that PE can not control the empirical sizes in most cases. However, the other methods can control the empirical sizes very well. So we do not consider the PE method in power comparison.

\begin{figure}[!ht]
\centering
\caption{Power of tests with different numbers of nonzero alpha at $T=100,N=200$ with normal errors.}
\includegraphics[width=6in,angle=0]{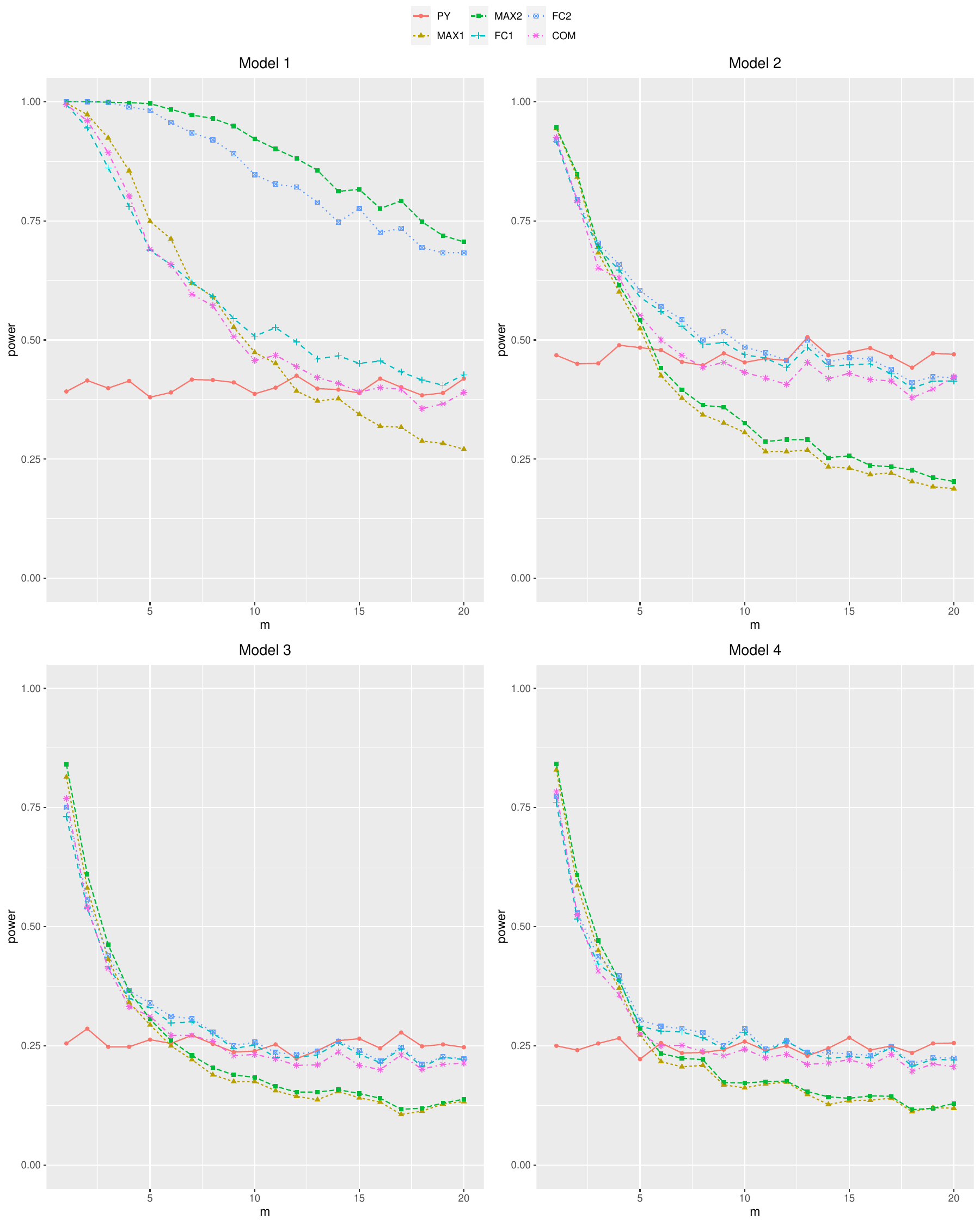}
\label{fig:pv1}
\end{figure}

\begin{figure}[!ht]
\centering
\caption{Power of tests with different numbers of nonzero alpha at $T=100,N=200$ with $t(5)$ errors.}
\includegraphics[width=6in,angle=0]{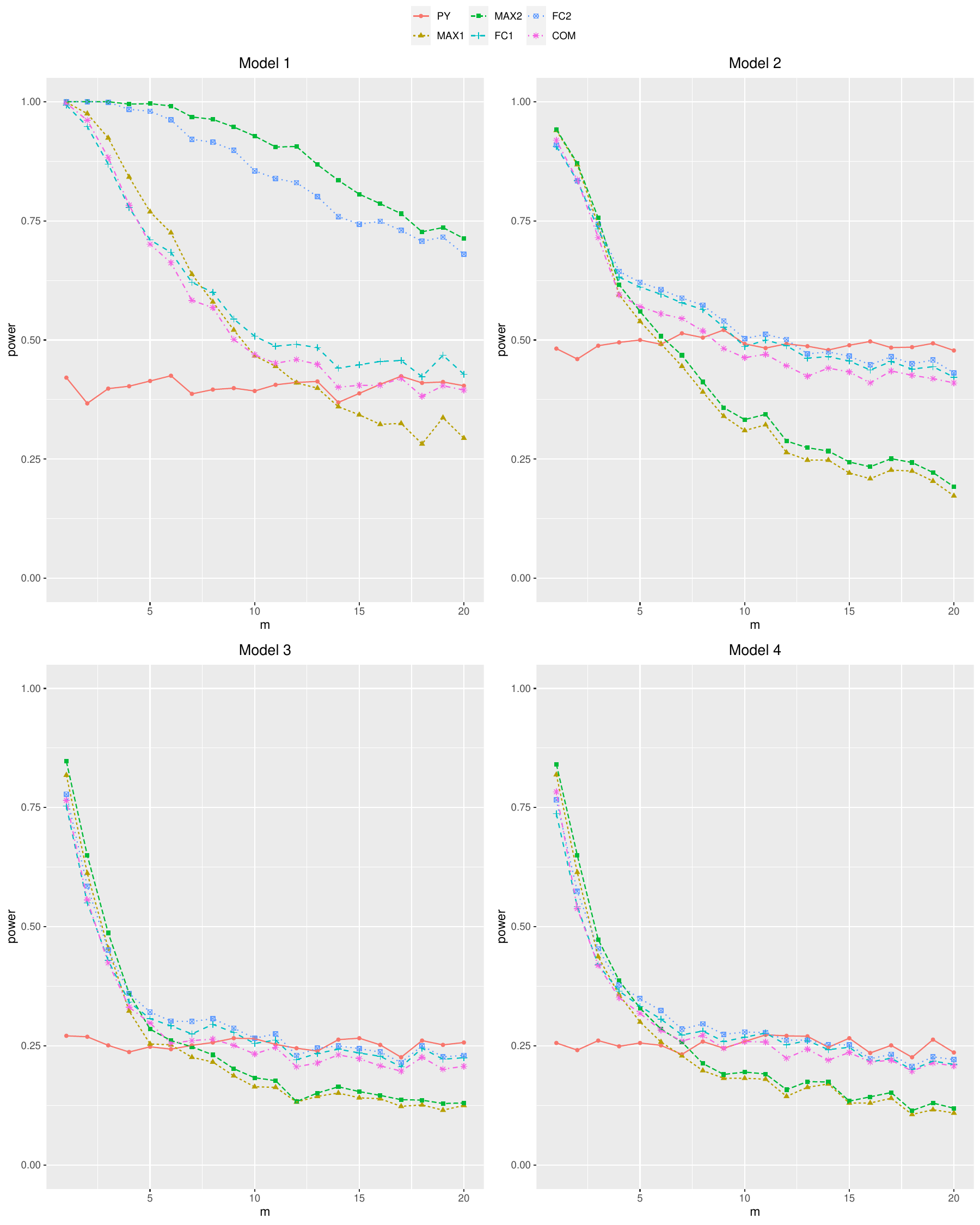}
\label{fig:pv2}
\end{figure}

\begin{figure}[!ht]
\centering
\caption{Power of tests with different numbers of nonzero alpha at $T=100,N=200$ with mixture normal errors.}
\includegraphics[width=6in,angle=0]{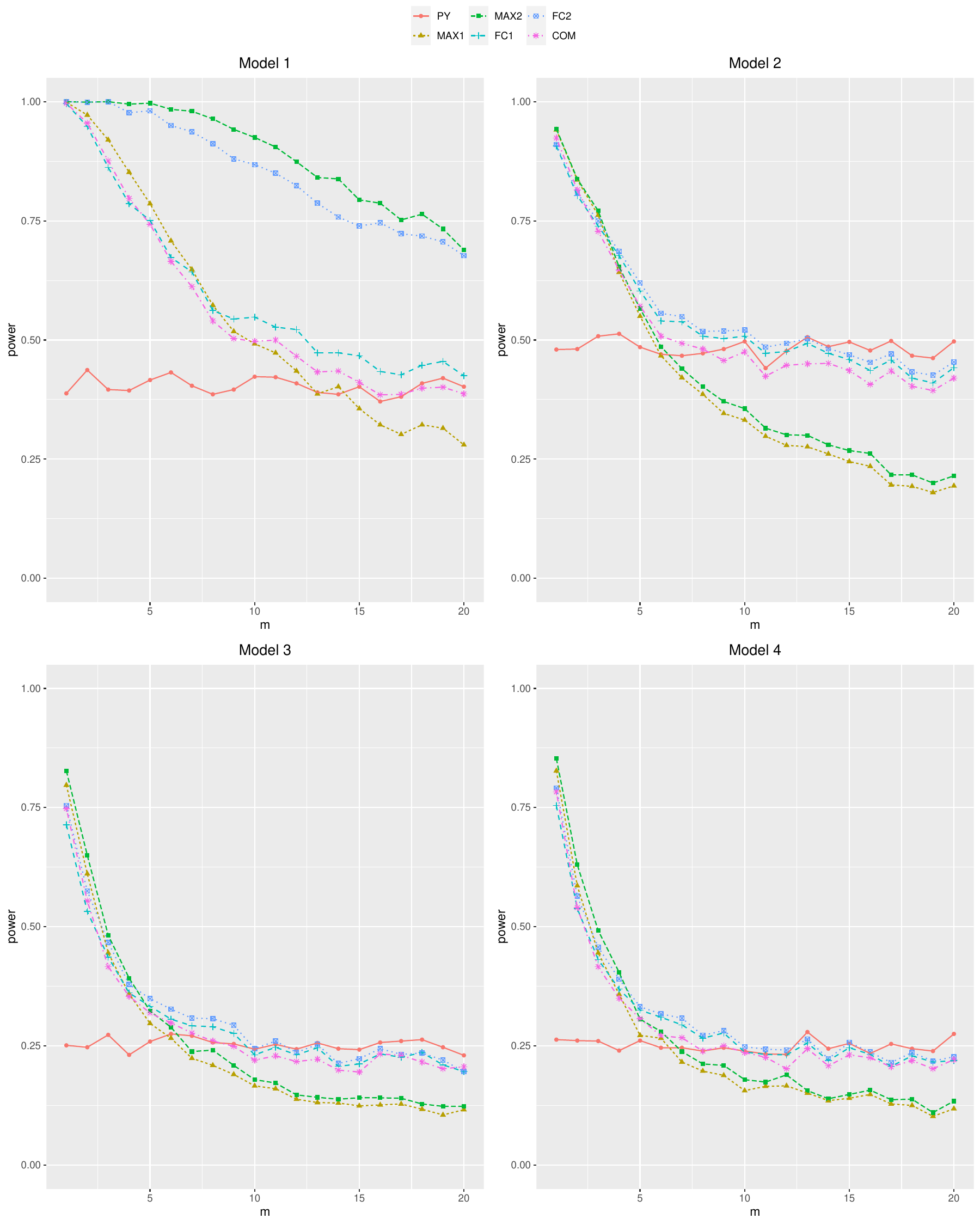}
\label{fig:pv3}
\end{figure}

To compare the power performance of the
various tests under different sparsity of $\bm \alpha$,
we present the empirical power of each test
under different number of nonzero elements of $\bm \alpha$. For illustration, here we only consider $T=100$, $N=200$. Specifically, given the number of nonzero elements of $\bm \alpha$,
i.e. $m\in \{1,\cdots,20\}$, we randomly choose a subset $S$ from
$\{1,\cdots,N\}$ with $|S|=m$
and let $\alpha_i=\sqrt{\frac{10\log(N)}{mT}}$ for $i \in S$,  such
that the signal strength $\|\bm \alpha\|^2$ is fixed for each $m$.
Figure \ref{fig:pv1}-\ref{fig:pv3} show the power curves of each tests under different error distributions.


In Model 2-4, the newly introduced max-type test, denoted as $M_{\hat{\O}^{1/2}}$, exhibits a performance that is comparable to $M_{\I}$, as referenced in \cite{feng2022high}. When we consider varying degrees of sparsity, it is observed that the sum-type test $T_{PY}$ surpasses the max-type tests $M_{\hat{\O}^{1/2}}$ and $M_{\I}$ as the value of $m$ increases. However, when $m$ is relatively small, the max-type tests outperform the sum-type test $T_{PY}$.

The combination tests demonstrate more consistent results across different levels of sparsity. Notably, when $m$ is neither too large nor too small, these combination tests prove to be more potent than both the max-type and sum-type tests. These observations align with our theoretical findings and corroborate many existing studies, such as \cite{feng2022asymptotic,feng2022high,chen2022asymptotic}.

Moreover, in Model 1, $M_{\hat{\O}^{1/2}}$ demonstrates significantly higher power than $M_{\I}$, underscoring the benefits of accounting for error correlation. As a result, the newly proposed adaptive test $T_{FC2}$ outperforms both $T_{FC1}$ and COM tests.

\section{Conclusion}
In this study, we initially introduce a novel max-type test statistic that takes into account the correlation among errors and exhibits superior performance under sparse alternatives. For broader alternatives, we suggest a Fisher combination test, which is predicated on the asymptotic independence between the newly proposed max-type test statistic and the sum-type test statistic from \cite{Pesaran2017Testing}. The efficacy of the proposed methods is further demonstrated through simulation studies.

Looking ahead, we propose several directions for future research. Firstly, the assumption of a constant regression coefficient may not hold in real-world scenarios. Therefore, numerous studies, such as \cite{ma2020testing} and \cite{ma2024adaptive}, have explored the testing of alpha in the high-dimensional time-varying factor model. It would be intriguing to propose a new max-type test and an adaptive test that standardizes the corresponding $t$-test statistic in the high-dimensional time-varying factor model. Secondly, the assumption of independence for the error term can be restrictive in practice. Thus, considering an adaptive test procedure with dependent observations, as suggested by \cite{ma2024testing}, poses a significant and interesting challenge. Finally, the independent component model, unfortunately, does not accommodate certain heavy-tailed distributions, such as the multivariate $t$-distribution. This is a limitation, given that financial data often exhibit heavy tails. Robust sum-type test procedures based on spatial signs for high-dimensional factor pricing models have been proposed by \cite{liu2023high}, \cite{zhao2022high}. The extension of these methods to our paper, particularly in conjunction with traditional spatial sign methods for sparse alternatives, is an area that warrants further exploration.

\section{Appendix}

\subsection{Proof of Theorem \ref{tho1}}
By the definition of $\hat\alpha_i$, we have $\hat\alpha_i=\sum_{t=1}^T\varepsilon_{it}c_t$ where $\bm c=(c_1,\cdots,c_T)^\top= (\bm 1_T^\top \P_{\X}\bm  1_T)^{-1}\P_{\X}\bm 1_T$. Let $\epsilon_{it}=\varepsilon_{it}/\sigma^{1/2}_{ii}$. We rewrite $$\tilde{t}=\sum_{t=1}^T \epsilon_{it}h_t$$ where $\bm h=(h_1,\cdots,h_T)^\top=(\bm 1_T^\top \P_{\X}\bm  1_T)^{-1/2}\P_{\X}\bm 1_T$. And we have $\sum_{t=1}^T h_t^2=1$.

Under condition (A1), we have $\O^{1/2}\bm{\tilde{t}} =\bm \upsilon=(\upsilon_1,\dots,\upsilon_N)$ where $\upsilon_i=\sum_{t=1}^T \xi_{it}h_t$.
Thus, taking the same procedure as Theorem 1 in \cite{feng2022high}, we have
\begin{align}\label{m1}
P\left(\max_{1\le i \le N}\vert\upsilon_i \vert \le \sqrt{2\log(N)-\log\log(N)+x} \right)\to \exp\{-\frac{1}{\sqrt{\pi}}\exp(-\frac{x}{2})\}
\end{align}

By the definition, we have $\hat{\O}^{1/2}\bm{{t}} =\bm \nu=(\nu_1,\dots,\nu_N)$ and
\begin{align*}
\left|||{\O}^{1/2}\bm t||_{\infty}-||{\O}^{1/2}\bm{\tilde{t}}||_{\infty}\right|\le ||{\O}^{1/2}(\bm t-\bm{\tilde{t})}||_{\infty}\le||\bm t-\bm{\tilde{t}}||_{\infty}||{\O}^{1/2}||_{L_1}
\end{align*}
\begin{align*}
\left|||\hat{\O}^{1/2}\bm t||_{\infty}-||{\O}^{1/2}\bm t||_{\infty}\right|\le ||(\hat{\O}^{1/2}-{\O}^{1/2})\bm t||_{\infty}\le ||{\O}^{1/2}\bm t||_{\infty} ||(\hat{\O}^{1/2}{\O}^{-1/2}-\I_N)||_{L_1}
\end{align*}
By the triangle inequality, we have
\begin{align*}
||\bm t-\bm{\tilde{t}}||_{\infty}=&\max_{1\le i\le N}\vert (\bm{t}-\bm{\tilde{t}})_i\vert = \max_{1\le i \le N}\vert \frac{\hat{\alpha}_i\sqrt{\bm 1_T^\top \P_{\X}\bm  1_T}}{\hat{\sigma}_{ii}^{1/2}}-\frac{\hat{\alpha}_i\sqrt{\bm 1_T^\top \P_{\X}\bm  1_T}}{{\sigma}_{ii}^{1/2}}\vert \\
\le& \max_{1 \leq i \leq N}\vert\frac{\hat{\alpha}_i\sqrt{\bm 1_T^\top \P_{\X}\bm  1_T}}{{\sigma}_{ii}^{1/2}}\vert \max_{1 \leq i \leq N}\vert \frac{\sigma_{ii}^{1/2}}{\hat{\sigma_{ii}}^{1/2}}-1\vert
\end{align*}
By Lemma E.2 and Proposition 4.1 in  \citet{Fan2015Power}, for some $c>0$, we have
\begin{align}
	P\{\max_{1 \leq i \leq N}\vert\hat{\sigma}_{i,i}-\sigma_{i,i}\vert>c\sqrt{\log(N)/T}\} \to 0
\end{align}
\begin{align}
	P\{\frac{4}{9}\le \frac{\hat{\sigma}_{i,i}}{\sigma_{i,i}}\le \frac{9}{4},i=1,\dots,N\} \to 1
\end{align}
Thus, with probability to one,we have
\begin{align*}
\max_{1 \leq i \leq N}\vert \frac{\sigma_{ii}^{1/2}}{\hat{\sigma_{ii}}^{1/2}}-1\vert \le&\max_{1 \leq i \leq N}\frac{1}{\hat{\sigma}_{ii}^{1/2}}\max_{1 \leq i \leq N}\vert\sigma_{ii}^{1/2}-\hat{\sigma}_{ii}^{1/2}\vert \\
\le& \frac{3}{2} \max_{1 \leq i \leq N}\frac{1}{\sigma_{ii}^{1/2}}\max_{1 \leq i \leq N}\vert\sigma_{ii}-\hat{\sigma}_{ii}\vert^{1/2} = O_p(\{\frac{log(N)}{T}\}^{1/4}).
\end{align*}
And by Theorem 1 in \cite{feng2022high}, we have
\begin{align*}
\max_{1 \leq i \leq N}\vert\frac{\hat{\alpha}_i\sqrt{\bm 1_T^\top \P_{\X}\bm  1_T}}{{\sigma}_i^{1/2}}\vert = O_p(\sqrt{\log(N)}),
\end{align*}
which leads to
\begin{align*}
\max_{1\le i \le N}\vert \frac{\hat{\alpha}_i\sqrt{\bm 1_T^\top \P_{\X}\bm  1_T}}{\hat{\sigma}_{ii}^{1/2}}-\frac{\hat{\alpha}_i\sqrt{\bm 1_T^\top \P_{\X}\bm  1_T}}{{\sigma}_{ii}^{1/2}}\vert = O_p({\frac{\{\log(N)\}^{3/4}}{T^{1/4}}}) \to 0
\end{align*}
due to $\log(N)=o(T^{1/4})$. So $||{\O}^{1/2}\bm t||_{\infty}-||{\O}^{1/2}\bm{\tilde{t}}||_{\infty}=o_p(1)$.
By (\ref{m1}), we have $||{\O}^{1/2}\bm t||_{\infty}=O_p(\{\log (N)\}^{1/2})$. By condition (A4), we have $||(\hat{\O}^{1/2}{\O}^{-1/2}-\I_N)||_{L_1}=o_p(\log^{-1}(N))$. So $||\hat{\O}^{1/2}\bm t||_{\infty}-||{\O}^{1/2}\bm t||_{\infty}=o_p(1)$.
Hence, we have
\begin{align*}
	P\left(\max_{1\le i \le N}\vert\nu_i \vert \le \sqrt{2\log(N)-\log\log(N)+x} \right)\to \exp\{-\frac{1}{\sqrt{\pi}}\exp(-\frac{x}{2})\}
\end{align*}
Thus,
\begin{align*}
	P\left(M_{{\hat{\O}}^{1/2}}-2\log(N)+\log\log (N)\le x\right)\to F(x)
\end{align*}
where \begin{align*}
	M_{\hat{{\O}}^{1/2}}=\max_{1\le i \le N} \nu_i^2.
\end{align*}
Here we obtain the result. \hfill$\Box$

\subsection{Proof of Theorem \ref{tho2}}
It suffices to prove
\begin{align*}
P\left(\max_{1 \leq i \leq N}\vert (\hat{\O}^{1/2}\bm{t})_i\vert \ge \sqrt{2\log(N)-\log\{\log(N)\}+q_\gamma} \right) \to 1.
\end{align*}
According to the proof of Theorem 1, we have
\begin{align*}
	P_{H_0}\left(\max_{1 \leq i \leq N}(\hat{\O}^{1/2}\bm{t})_i^2-2\log(N)+\log\{\log(N)\}\le x\right) \to \exp\{-\frac{1}{\sqrt{\pi}}\exp(-\frac{x}{2})\}.
\end{align*}
By lemma 3 in \cite{Cai2014}, for $k_N$-sparse $\bm{t}$, $r<\frac{1}{4}$, we have, for any $2r<a<1-2r$,as $p \to \infty$
\begin{align*}
	P\left( \max_{i \in H} \vert(\O^{1/2}\bm{{t}})_i - \omega_{i,i}t_i \vert = O_p(N^{r-a/2})\max_{i \in H}\vert t_i \vert \right) \to 1
\end{align*}
where $H$ is the support of $\bm{t}$.  And as $||\hat{\O}^{1/2}\bm t||_{\infty}-||{\O}^{1/2}\bm{{t}}||_{\infty}=o_p(1)$,  we have
\begin{align*}
	P_{H_0}\left(\max_{1 \leq i \leq N}({\omega}_{i, i}t_i)^2-2\log(N)+\log\{\log(N)\}\le x\right) \to \exp\{-\frac{1}{\sqrt{\pi}}\exp(-\frac{x}{2})\}.
\end{align*}
Thus,
\begin{align*}
	P\left(\max_{1 \leq i \leq N}{\omega}_{i,i}^2 \frac{(\hat{\alpha}_i-\alpha_i)^2\bm 1_T^\top \P_{\X}\bm  1_T}{\hat{\sigma}_{i,i}} \le 2\log(N)-\frac{1}{2}\log\{\log(N)\}\right) \to 1
\end{align*}
when we set $x=\frac{1}{2}\log{\log(N)}$.
By the triangle inequality, we have
\begin{align*}
	&\max_{1 \leq i \leq N}{\omega}_{i,i}^2 \frac{\bm 1_T^\top \P_{\X}\bm  1_T{\hat{\alpha}_i}^2}{\hat{\sigma}_{i,i}} \\
	\ge & \max_{1 \leq i \leq N}{\omega}_{i,i}^2 \frac{\bm 1_T^\top \P_{\X}\bm  1_T{{\alpha}_i}^2}{2\hat{\sigma}_{i,i}}- \max_{1 \leq i \leq N}{\omega}_{i,i}^2 \frac{\bm 1_T^\top \P_{\X}\bm  1_T{(\hat{\alpha}_i-\alpha_i)}^2}{\hat{\sigma}_{i,i}} \\
	\ge &  \max_{1 \leq i \leq N}{\omega}_{i,i}^2 \frac{\bm 1_T^\top \P_{\X}\bm  1_T{{\alpha}_i}^2}{2{\sigma}_{i,i}}- \max_{1 \leq i \leq N}{\omega}_{i,i}^2 \frac{\bm 1_T^\top \P_{\X}\bm  1_T{(\hat{\alpha}_i-\alpha_i)}^2}{\hat{\sigma}_{i,i}} - \sqrt{\frac{\log(N)}{T}}\max_{1 \leq i \leq N}{\omega}_{i,i}^2 \frac{\bm 1_T^\top \P_{\X}\bm  1_T{{\alpha}_i}^2}{2{\sigma}_{i,i}}\\
	\ge & 4\log(N)-2\log(N)+\frac{1}{2}\log{\log(N)}-\sqrt{\frac{\log(N)}{T}}\max_{1 \leq i \leq N}{\omega}_{i,i}^2 \frac{\bm 1_T^\top \P_{\X}\bm  1_T{{\alpha}_i}^2}{2{\sigma}_{i,i}}
\end{align*}
with probability tending to 1. If $\max_{1 \leq i<j \leq N}{\omega}_{i,i}^2 \frac{\bm 1_T^\top \P_{\X}\bm  1_T{{\alpha}_i}^2}{2{\sigma}_{i,i}}=O\{\log(N)\}$
\begin{align*}
\max_{1 \leq i \leq N}{\omega}_{i,i}^2 \frac{\bm 1_T^\top \P_{\X}\bm  1_T{{\alpha}_i}^2}{2{\sigma}_{i,i}}	\ge & 4\log(N)-2\log(N)+\frac{1}{2}\log\{\log(N)\}-O\left(\sqrt{\frac{\{\log(N)\}^3}{T}}\right) \\
\ge & 2\log(N)-\log\{\log(N)\}+q_\gamma
\end{align*}
which implies $P(\phi_{\gamma}=1) \to 1$. If ${\log^{-1}{(N)}}\max_{1 \leq i \leq N}{\omega}_{i,i}^2 \frac{\bm 1_T^\top \P_{\X}\bm  1_T{{\alpha}_i}^2}{2{\sigma}_{i,i}} \to \infty$
\begin{align*}
\max_{1 \leq i \leq N}{\omega}_{i,i}^2 \frac{\bm 1_T^\top \P_{\X}\bm  1_T{\hat{\alpha}_i}^2}{\hat{\sigma}_{i,i}} \ge \frac{4}{9}\max_{1 \leq i \leq N}{\omega}_{i,i}^2 \frac{\bm 1_T^\top \P_{\X}\bm  1_T{\hat{\alpha}_i}^2}{{\sigma}_{i,i}} \ge 2\log(N)-\log\{\log(N)\}+q_\gamma
\end{align*}
Thus, $P(\phi_{\gamma}=1)$ converges to one.
\subsection{Some useful Theorems and Lemmas}

We also restate Lemma S.10 in \cite{feng2022asymptotic} here.
\begin{lemma}\lbl{small_lemma} Let $\{(U, U_{N},\tilde{U}_N)\in \mathbb{R}^3;\, N\geq 1\}$ and  $\{(V, V_{N},\tilde{V}_N)\in \mathbb{R}^3;\, N\geq 1\}$ be two sequences of random variables with $U_N\to U$ and $V_N\to V$ in distribution as $N\to\infty.$ Assume $U$ and $V$ are continuous random variables. We assume that
\begin{align}\label{aixiaokey}
\tilde{U}_N=U_N+o_p(1),\tilde{V}_N=V_N+o_p(1).
\end{align} If $U_N$ and $V_N$ are asymptotically independent, then $\tilde{U}_N$ and
$\tilde V_N$ are also asymptotically independent.
\end{lemma}

Define $B_i=\{\upsilon_i^2>l_N\}$ and $A_N(x)=\left\{\frac{\bm \upsilon^\top \R \bm \upsilon-N}{\sqrt{2\tr(\R^2)}}<x\right\}$. Define $l_N=2\log(N)-2\log\log(N)+y$. We first proof the following important lemma. Define $h(y)=\frac{1}{\sqrt{\pi}}e^{-y/2}$.
\begin{lemma}\label{le1}
Under the assumption of Theorem \ref{tho3}, for each $d\ge 1$, we have
\begin{align}\label{le11}
\lim_{N\to \infty} H(d,N) \le \frac{1}{d!} h^d(y)<\infty
\end{align}
where $H(d,N)\doteq \sum_{1\le i_1<\cdots<i_d\le N} P(B_{i_1}\cdots B_{i_d})$.
And then, we have
\begin{equation}\label{le12}
\sum_{1 \leq i_{1}<\cdots<i_{d} \leq N}\left|P\left(A_{N}(x) B_{i_{1}} \cdots B_{i_{d}}\right)-P\left(A_{N}(x)\right) \cdot P\left(B_{i_{1}} \cdots B_{i_{d}}\right)\right| \rightarrow 0
\end{equation}
as $N\to \infty$.
\end{lemma}
\proof  According to the proof of Theorem 1 in \cite{feng2022high}, we have
\begin{align*}
NP(\upsilon_i^2\ge l_N)=2N(1-\Phi(\sqrt{l_N}))+o(1)=\frac{1}{\sqrt{\pi}}e^{-y/2}+o(1)
\end{align*}
as $T,N\to \infty$ and $N/T^2\to 0$.
 Because $NP(B_i)\to h(y)$, we have $NB(B_i)<h(y)+\epsilon$ for any $\epsilon>0$ as $N\to \infty$. By the independence of $z_i$, we have
\begin{align*}
H(d,N)=&\sum_{1\le i_1<\cdots<i_d\le N} P(B_{i_1}\cdots B_{i_d})=\sum_{1\le i_1<\cdots<i_d\le N} \prod_{k=1}^d P(B_{i_k})\\
\le& C_N^d \{N^{-1}(h(y)+\epsilon)\}^d\le \frac{1}{d!} \left(h(y)+\epsilon\right)^d
\end{align*}
So, by letting $\epsilon \to 0$, we have
\begin{align*}
\lim_{N\to \infty}H(d,N) \le \frac{1}{d!} h^d(y)<\infty.
\end{align*}
Here we prove (\ref{le11}).

Additionally, because of $e^x\le 1+(1+\epsilon)|x|$ for $x<\epsilon$, we have
\begin{align*}
E(\exp(\lambda \upsilon_i))&=\prod_{t=1}^T E(e^{\lambda \xi_{it}h_t}) \le \prod_{t=1}^T (1+(1+\epsilon)\lambda E( |\xi_{it}h_t|)))\\
&\le (1+(1+\epsilon)C_hK^{1/(8+c)}T^{-1/2}\lambda ))^T\\
&\le \exp((1+\epsilon)^2C^2_hK^{2/(8+c)}\lambda^2)
\end{align*}
Thus, we have $\upsilon_i$ is a sub-gaussian random variable. So there exist $\eta>0$ and $K>0$ such that $E(\exp(\eta \upsilon_i^2))\le K$.

Define $\bu=(\bu_1,\bu_2)$ where $\bu_1=(\upsilon_1,\cdots,\upsilon_d)$ and $\bu_2=(\upsilon_{d+1},\cdots,\upsilon_N)$. And
\begin{align*}
\R=\left(
\begin{array}{cc}
\R_{11}&\R_{12}\\
\R_{21}&\R_{22}
\end{array}
\right)
\end{align*}
So,
\begin{align*}
\bu^\top\R\bu=\bu_{1}^\top \R_{11}\bu_1+2\bu_1^\top \R_{12}\bu_2+\bu_2^\top \R_{22}\bu_2.
\end{align*}

Because $\lambda_{\max}(\R_{11})\le \lambda_{\max}(\R)<c$,
\begin{align*}
P\left(\bu_{1}^\top \R_{11}\bu_1>\epsilon\sqrt{2\tr(\R^2)}\right)\le& P\left(c\bu_{1}^\top \bu_1>\epsilon\sqrt{2\tr(\R^2)}\right)\\
= & P\left(\eta\sum_{i=1}^d z_i^2>c^{-1}\eta\epsilon\sqrt{2\tr(\R^2)}\right)\\
\le & \exp\left(-c^{-1}\eta\epsilon\sqrt{2\tr(\R^2)}\right) E(e^{\eta\sum_{i=1}^d z_i^2})\\
=&\exp\left(-c^{-1}\eta\epsilon\sqrt{2\tr(\R^2)}\right) \{E(e^{\eta z_i^2})\}^d\\
\le & K^{d}\exp\left(-c^{-1}\eta\epsilon\sqrt{2\tr(\R^2)}\right)
\end{align*}
By the assumption (iii), we have $\sqrt{2\tr(\R^2)}^2\ge 2\tr(\R^2)\ge 2c^{-2}N$. So
\begin{align}\label{s1}
P\left(\bu_{1}^\top \R_{11}\bu_1>\epsilon\sqrt{2\tr(\R^2)}\right)\le K^{d}\exp\left(-\sqrt{2}c^{-2}\eta\epsilon N^{1/2}\right).
\end{align}
Define $\mathbf{A}=\mathbf{Q}^{\top} \boldsymbol{\Lambda} \mathbf{Q}$ where $\mathbf{Q}=\left(q_{i j}\right)_{1 \leq i, j \leq p}$ is an orthogonal matrix and $\boldsymbol{\Lambda}=\operatorname{diag}\left\{\lambda_{1}, \ldots, \lambda_{N}\right\}, \lambda_{i}, i=1, \ldots, N$ are the eigenvalues of $\mathbf{A}$. Note that $\sum_{1 \leq j \leq N} a_{i j}^{2}$ is the $i$ th diagonal element of $\mathbf{A}^{2}=\mathbf{Q}^{\top} \mathbf{\Lambda}^{2} \mathbf{Q}$, we have $\sum_{1 \leq j \leq N} a_{i j}^{2}=\sum_{l=1}^{N} q_{l i}^{2} \lambda_{l}^{2} \leq c^{2}$ according to Assumption (ii).

Next, define $\theta=\sqrt{\frac{2\eta}{dc^2\sigma^2}}$, we have
\begin{align*}
P\left(\bu_1^\top \R_{12}\bu_2\ge \epsilon\sqrt{2\tr(\R^2)}\right)\le& \exp\left(-\theta\epsilon\sqrt{2\tr(\R^2)}\right) E\left(\exp(\theta\bu_1^\top \R_{12}\bu_2)\right)\\
=&\exp\left(-\theta\epsilon\sqrt{2\tr(\R^2)}\right) E(e^{\theta\sum_{i=1}^d\sum_{j=d+1}^Na_{ij}z_iz_j})\\
\le &\exp\left(-\theta\epsilon\sqrt{2\tr(\R^2)}\right) E(E(e^{\theta\sum_{j=d+1}^N(\sum_{i=1}^da_{ij}z_i)z_j}|\bu_1))\\
=&\exp\left(-\theta\epsilon\sqrt{2\tr(\R^2)}\right) E\left(\prod_{j=d+1}^{N} E(e^{(\theta\sum_{i=1}^da_{ij}z_i)z_j}|\bu_1)\right)\\
\le &\exp\left(-\theta\epsilon\sqrt{2\tr(\R^2)}\right) E\left(\prod_{j=d+1}^{N} \exp\left(\frac{\sigma^2\theta^2}{2}\left(\sum_{i=1}^da_{ij}z_i\right)^2\right)\right)\\
=&\exp\left(-\theta\epsilon\sqrt{2\tr(\R^2)}\right) E\left( \exp\left(\frac{\sigma^2\theta^2}{2}\sum_{j=d+1}^{N} \left(\sum_{i=1}^da_{ij}z_i\right)^2\right)\right)\\
\le &\exp\left(-\theta\epsilon\sqrt{2\tr(\R^2)}\right) E\left( \exp\left(\frac{d\sigma^2\theta^2}{2}\sum_{j=d+1}^{N}\sum_{i=1}^da_{ij}^2z_i^2\right)\right)\\
\le &\exp\left(-\theta\epsilon\sqrt{2\tr(\R^2)}\right) E\left( \exp\left(\frac{dc^2\sigma^2\theta^2}{2}\sum_{i=1}^dz_i^2\right)\right)\\
= &\exp\left(-\theta\epsilon\sqrt{2\tr(\R^2)}\right) E\left( \exp\left(\eta\sum_{i=1}^dz_i^2\right)\right)\\
\le &K^d\exp\left(-\theta\epsilon\sqrt{2\tr(\R^2)}\right)\le K^d \exp\left(-\sqrt{2}c^{-1}\theta\epsilon N^{1/2}\right)
\end{align*}
So
\begin{align}\label{s2}
P\left(\bu_1^\top \R_{12}\bu_2\ge \epsilon\sqrt{2\tr(\R^2)}\right)\le&K^d \exp\left(-\sqrt{\frac{4\eta}{dc^4\sigma^2}}\epsilon N^{1/2}\right)
\end{align}
Similarly, we also can prove that
\begin{align}\label{s3}
P\left((-\bu_1)^\top \R_{12}\bu_2\ge \epsilon\sqrt{2\tr(\R^2)}\right)\le&K^d \exp\left(-\sqrt{\frac{4\eta}{dc^4\sigma^2}}\epsilon N^{1/2}\right)
\end{align}
Let $\Theta_p=\bu_{1}^\top \R_{11}\bu_1+2\bu_1^\top \R_{12}\bu_2$.
\begin{align*}
&P\left(|\Theta_N|>\epsilon\sqrt{2\tr(\R^2)}\right)\\
\le& P\left(\bu_{1}^\top \R_{11}\bu_1>\epsilon\sqrt{2\tr(\R^2)}/2\right)+P\left(|\bu_1^\top \R_{12}\bu_2|>\epsilon\sqrt{2\tr(\R^2)}/4\right)\\
\le &P\left(\bu_{1}^\top \R_{11}\bu_1>\epsilon\sqrt{2\tr(\R^2)}/2\right)+P\left(\bu_1^\top \R_{12}\bu_2>\epsilon\sqrt{2\tr(\R^2)}/8\right)+P\left(-\bu_1^\top \R_{12}\bu_2>\epsilon\sqrt{2\tr(\R^2)}/8\right)\\
\end{align*}
So, by (\ref{s1}), (\ref{s2}) and (\ref{s3}), there exist a constant $c_{\epsilon}>0$,
\begin{align}\label{s4}
P\left(|\Theta_N|>\epsilon\sqrt{2\tr(\R^2)}\right)\le K^d\exp(-c_{\epsilon}N^{1/2})
\end{align}
\begin{align*}
&P\left(A_{N}(x) B_{1} \cdots B_{d}\right)\\
=&P\left(\frac{\bu_2^\top \R_{22}\bu_2-\tr(\R)+\Theta_N}{\sqrt{2\tr(\R^2)}}\le x, B_1\cdots B_d\right)\\
\le &P\left(\frac{\bu_2^\top \R_{22}\bu_2-\tr(\R)+\Theta_N}{\sqrt{2\tr(\R^2)}}\le x, |\Theta_N|\le\epsilon \sqrt{2\tr(\R^2)}, B_1\cdots B_d\right)+P\left(|\Theta_N|>\epsilon \sqrt{2\tr(\R^2)}\right)\\
\le &P\left(\frac{\bu_2^\top \R_{22}\bu_2-\tr(\R)}{\sqrt{2\tr(\R^2)}}\le x+\epsilon, B_1\cdots B_d\right)+K^d\exp(-c_{\epsilon}N^{1/2})\\
=&P\left(\frac{\bu_2^\top \R_{22}\bu_2-\tr(\R)}{\sqrt{2\tr(\R^2)}}\le x+\epsilon\right)P\left(B_1\cdots B_d\right)+K^d\exp(-c_{\epsilon}N^{1/2})\\
\le &\left[P\left(\frac{\bu_2^\top \R_{22}\bu_2-\tr(\R)+\Theta_N}{\sqrt{2\tr(\R^2)}}\le x+\epsilon,|\Theta_N|\le\epsilon \sqrt{2\tr(\R^2)}\right)+P\left(|\Theta_N|>\epsilon \sqrt{2\tr(\R^2)}\right)\right]P\left(B_1\cdots B_d\right)\\
&+K^d\exp(-c_{\epsilon}N^{1/2})\\
\le & P\left(\frac{\bu_2^\top \R_{22}\bu_2-\tr(\R)+\Theta_n}{\sqrt{2\tr(\R^2)}}\le x+2\epsilon\right)P\left(B_1\cdots B_d\right)+2K^d\exp(-c_{\epsilon}N^{1/2})\\
=&P\left(A_p( x+2\epsilon)\right)P\left(B_1\cdots B_d\right)+2K^d\exp(-c_{\epsilon}N^{1/2})
\end{align*}
Similarly, we can prove that
\begin{align*}
P\left(A_{N}(x) B_{1} \cdots B_{d}\right)\ge P\left(A_N( x-2\epsilon)\right)P\left(B_1\cdots B_d\right)-2K^d\exp(-c_{\epsilon}N^{1/2})
\end{align*}
So, we have
\begin{align}\label{pba}
\left|P\left(A_{N}(x) B_{1} \cdots B_{d}\right)-P\left(A_{p}(x)\right) \cdot P\left(B_{1} \cdots B_{d}\right)\right|
\leq  \Delta_{N, \epsilon} \cdot P\left(B_{1} \cdots B_{d}\right)+2K^d\exp(-c_{\epsilon}N^{1/2})
\end{align}
where
\begin{align*}
\Delta_{p, \epsilon}&=\left|P\left(A_{N}(x)\right)-P\left(A_{N}(x+2 \epsilon)\right)\right|+\left|P\left(A_{N}(x)\right)-P\left(A_{N}(x-2 \epsilon)\right)\right| \\
&=P\left(A_{N}(x+2 \epsilon)\right)-P\left(A_{N}(x-2 \epsilon)\right)
\end{align*}
Obviously, the inequality (\ref{pba}) holds for all $i_1,\cdots,i_d$. Thus,
\begin{align*}
&\sum_{1 \leq i_{1}<\cdots<i_{d} \leq p}\left|P\left(A_{N}(x) B_{i_{1}} \cdots B_{i_{d}}\right)-P\left(A_{N}(x)\right) \cdot P\left(B_{i_{1}} \cdots B_{i_{d}}\right)\right| \\
&\leq \sum_{1 \leq i_{1}<\cdots<i_{d} \leq p}\left[\Delta_{N, \epsilon} \cdot P\left(B_{i_{1}} \cdots B_{i_{d}}\right)+2K^d\exp(-c_{\epsilon}N^{1/2})\right] \\
&\leq \Delta_{N, \epsilon} \cdot H(d,N)+\left(\begin{array}{l}
N \\
d
\end{array}\right) \cdot 2K^d\exp(-c_{\epsilon}N^{1/2})
\end{align*}
By Theorem 3 in Pesaran and Yamagata (2021), we have $P(A_N(x))\to \Phi(x)$ as $p\to \infty$. So $\Delta_{N, \epsilon}\to P(x+2\epsilon)-P(x-2\epsilon)$. By letting $\epsilon \to 0$, we have $\Delta_{p, \epsilon}\to 0$. By (\ref{le11}), we have $\lim_{p\to \infty}H(d,p)<\infty$. Additionally, $\left(\begin{array}{l}
N \\
d
\end{array}\right) \cdot 2K^d\exp(-c_{\epsilon}N^{1/2})\to 0$ as $N\to\infty$. So we can obtain (\ref{le12}). \hfill$\Box$
\subsection{Proof of Theorem \ref{tho3}}
\paragraph{Proof of Theorem \ref{tho3}}

According to the proof of Theorem \ref{tho1}, we have
\begin{align*}
\max_{1\le i\le N} \nu_i^2-\max_{1\le i\le N} \upsilon_i^2=o_p(1)
\end{align*}
where
\begin{align}
\upsilon_i=\sum_{t=1}^T \xi_{it}h_t.
\end{align}
And we have $\sum_{i=1}^N z_i^2=\bm \upsilon^\top \R \bm \upsilon$ where $\bm \upsilon=(\upsilon_1,\cdots,\upsilon_N)$.
Thus, according to Lemma \ref{small_lemma}, we only need to show that
\begin{align}
P\left(\frac{\bm \upsilon^\top \R \bm \upsilon-N}{\sqrt{2\tr(\R^2)}}<x,\max_{1\le i\le N} \upsilon_i^2-2\log N+\log\log N<y\right)\to \Phi(x)F(y)
\end{align}

 Additionally, by Theorem 3 in Pesaran and Yamagata (2021) we know that
\begin{align}\label{i3}
P\left(\frac{\bu^\top\R\bu-\tr(\R)}{\sqrt{2\tr(\R^2)}}\le x\right)\to \Phi(x)
\end{align}
To show (\ref{asy2}), we only need to show that
\begin{align}\label{th12}
P\left(\frac{\bu^\top\R\bu-\tr(\R)}{\sqrt{2\tr(\R^2)}}\le x,\max_{1\le i\le N}\upsilon^2_i> l_N(y)\right)\to\Phi(x)(1-F(y))
\end{align}
Recall the notations in Lemma \ref{le1}, we have
\bea\lbl{abci}
P\Big(\frac{\bu^\top\R\bu-\tr(\R)}{\sqrt{2\tr(\R^2)}}\leq x,\ \max_{1\le i\le N}\upsilon^2_i>l_N\Big)=P\Big(\bigcup_{i=1}^NA_NB_{i}\Big).
\eea
Here the notation $A_NB_i$ stands for $A_N\cap B_i$ and we brief $A_N(x)$ as $A_p$.  From the inclusion-exclusion principle,
\bea
P\Big(\bigcup_{i=1}^NA_NB_{i}\Big)  \leq  \sum_{1\leq i_1 \leq N}P(A_NB_{i_1})&-&\sum_{1\leq i_1< i_2\leq N}P(A_NB_{i_1}B_{i_2})+\cdots+\nonumber\\
& & \sum_{1\leq i_1<  \cdots < i_{2k+1}\leq N}P(A_NB_{i_1}\cdots B_{i_{2k+1}})\nonumber\\
&&  \lbl{Upper_bound}
\eea
and
\bea
P\Big(\bigcup_{i=1}^NA_NB_{i}\Big)  \geq  \sum_{1\leq i_1 \leq N}P(A_NB_{i_1})&-&\sum_{1\leq i_1< i_2\leq N}P(A_NB_{i_1}B_{i_2})+\cdots- \nonumber\\
& & \sum_{1\leq i_1<  \cdots < i_{2k}\leq  N}P(A_NB_{i_1}\cdots B_{i_{2k}}) \nonumber\\
&&  \lbl{Lower_bound}
\eea
for any integer $k\geq 1$. Define
\beaa
H(N, d)=\sum_{1\leq i_1<  \cdots < i_{d}\leq p}P(B_{i_1}\cdots B_{i_{d}})
\eeaa
for $d\geq 1$. From \eqref{le11} we know
%
%
\bea\lbl{Maya1}
\lim_{d\to\infty}\limsup_{N\to\infty}H(p, d)=0.
\eea
Set
\beaa
\zeta(N,d)=\sum_{1\leq i_1<  \cdots < i_d\leq N}\big[P(A_NB_{i_1}\cdots B_{i_d}) - P(A_N)\cdot P(B_{i_1}\cdots B_{i_d})\big]
\eeaa
for $d\geq 1.$ By Lemma \ref{le1},
\bea\lbl{back_campus}
\lim_{N\to\infty}\zeta(N,d)=0
\eea
for each $d\geq 1$. The assertion \eqref{Upper_bound} implies that
\bea\lbl{639475}
P\Big(\bigcup_{i=1}^NA_NB_{i}\Big)
& \leq & P(A_N)\Big[\sum_{1\leq i_1 \leq N}P(B_{i_1})-\sum_{1\leq i_1< i_2\leq N}P(B_{i_1}B_{i_2})+\cdots-  \nonumber\\
&& \sum_{1\leq i_1<  \cdots < i_{2k} \leq N}P(B_{i_1}\cdots B_{i_{2k}})\Big]+ \Big[\sum_{d=1}^{2k}\zeta(N,d)\Big] + H(N, 2k+1)  \nonumber\\
&\leq & P(A_N)\cdot P\Big(\bigcup_{i=1}^NB_{i}\Big)+ \Big[\sum_{d=1}^{2k}\zeta(N,d)\Big] + H(N, 2k+1),
\eea
where the inclusion-exclusion formula is used again in the last inequality, that is,
\beaa
P\Big(\bigcup_{i=1}^NB_{i}\Big) &\geq & \sum_{1\leq i_1 \leq N}P(B_{i_1})-\sum_{1\leq i_1< i_2\leq N}P(B_{i_1}B_{i_2})+\cdots - \nonumber\\
&&~~~~~~~~~~~~~~~~ \sum_{1\leq i_1<  \cdots < i_{2k}\leq N}P(B_{i_1}\cdots B_{i_{2k}})
\eeaa
for all $k\geq 1$.
 By the definition of $l_N$ and Theorem \ref{tho1},
\beaa
 P\Big(\bigcup_{i=1}^NB_{i}\Big) \to 1-F(y)
\eeaa
as $N\to\infty$. By \eqref{i3}, $P(A_N)\to \Phi(x)$ as $N\to\infty.$ From \eqref{abci}, \eqref{back_campus} and \eqref{639475}, by fixing $k$ first and sending $N\to \infty$ we obtain that
\beaa
\limsup_{N\to\infty}P\Big(\frac{\bu^\top\R\bu-\tr(\R)}{\sqrt{2\tr(\R^2)}}\leq x,\ \max_{1\le i\le N}\upsilon^2_i>l_N\Big)\leq \Phi(x)\cdot [1-F(y)] +\lim_{N\to\infty}H(N, 2k+1).
\eeaa
Now, by letting $k\to \infty$ and using \eqref{Maya1} we have
\bea\lbl{vskdnti}
\limsup_{N\to\infty}P\Big(\frac{\bu^\top\R\bu-\tr(\R)}{\sqrt{2\tr(\R^2)}}\leq x,\ \max_{1\le i\le N}\upsilon^2_i>l_N\Big)\leq \Phi(x)\cdot [1-F(y)].
\eea
By applying the same argument to \eqref{Lower_bound}, we see that the counterpart of \eqref{639475} becomes
\beaa
P\Big(\bigcup_{i=1}^NA_NB_{i}\Big)
& \geq & P(A_N)\Big[\sum_{1\leq i_1 \leq N}P(B_{i_1})-\sum_{1\leq i_1< i_2\leq N}P(B_{i_1}B_{i_2})+\cdots + \nonumber\\
&& \sum_{1\leq i_1<  \cdots < i_{2k-1}\leq N}P(B_{i_1}\cdots B_{i_{2k-1}})\Big] + \Big[\sum_{d=1}^{2k-1}\zeta(N,d)\Big] - H(N, 2k)  \nonumber\\
&\geq & P(A_N)\cdot P\Big(\bigcup_{i=1}^NB_{i}\Big) + \Big[\sum_{d=1}^{2k-1}\zeta(N,d)\Big] - H(N, 2k).
\eeaa
where in the last step we use the inclusion-exclusion principle such that
\beaa
P\Big(\bigcup_{i=1}^NB_{i}\Big) &\leq & \sum_{1\leq i_1 \leq N}P(B_{i_1})-\sum_{1\leq i_1< i_2\leq N}P(B_{i_1}B_{i_2})+\cdots + \nonumber\\
&&~~~~~~~~~~~~~~~~ \sum_{1\leq i_1<  \cdots < i_{2k-1}\leq N}P(B_{i_1}\cdots B_{i_{2k-1}})
\eeaa
for all $k\geq 1$. Review \eqref{abci} and repeat the earlier procedure to see
\beaa
\liminf_{N\to\infty}P\Big(\frac{\bu^\top\R\bu-\tr(\R)}{\sqrt{2\tr(\R^2)}}\leq x,\ \max_{1\le i\le N}\upsilon^2_i>l_N\Big)\geq \Phi(x)\cdot [1-F(y)]
\eeaa
by sending $N\to \infty$ and then sending $k\to\infty.$
This and \eqref{vskdnti} yield \eqref{th12}. The proof is completed. \hfill$\Box$

\subsection{Proof of Theorem \ref{tho33}}
According to the proof of Theorem \ref{tho3}, we only need to show that
\begin{align*}
&P\left(\frac{\tilde{\bm \upsilon}^\top \R \tilde{\bm \upsilon}-N}{\sqrt{2\tr(\R^2)}}<x,\max_{1\le i\le N} \tilde\upsilon_i^2-2\log N+\log\log N<y\right)\\
&\to P\left(\frac{\tilde{\bm \upsilon}^\top \R\tilde{ \bm \upsilon}-N}{\sqrt{2\tr(\R^2)}}<x\right)P\left(\max_{1\le i\le N}\tilde\upsilon_i^2-2\log N+\log\log N<y\right)
\end{align*}
where $\tilde \upsilon_i=\upsilon_i+\tilde{\alpha}_i\sum_{t=1}^T h_t$. Define $\bm \omega=(\omega_1,\cdots,\omega_N)^\top$ and $\omega_i=\tilde{\alpha}_i\sum_{t=1}^T h_t=O(T^{1/2}\alpha_i)$. We have
\begin{align*}
\max_{1\le i\le N} \tilde \upsilon_i^2=\max\{\max_{i \in \mathcal{M}} \tilde\upsilon_i^2, \max_{i \in \mathcal{M}^c} \tilde\upsilon_i^2\}\doteq \max\{M_1,M_2\}
\end{align*}
and
\begin{align*}
\frac{\tilde{\bm \upsilon}^\top \R \tilde{\bm \upsilon}-N}{\sqrt{2\tr(\R^2)}}=&\frac{\bm \upsilon_{\mathcal{M}^c}^\top \R_{\mathcal{M}^c} \bm \upsilon_{\mathcal{M}^c}-N}{\sqrt{2\tr(\R^2)}}+\frac{2\bm \upsilon_{\mathcal{M}^c}^\top \R_{\mathcal{M}^c\mathcal{M}} \tilde{\bm \upsilon}_{\mathcal{M}}}{\sqrt{2\tr(\R^2)}}+\frac{\tilde{\bm \upsilon}_{\mathcal{M}}^\top \R_{\mathcal{M}} \tilde{\bm \upsilon}_{\mathcal{M}}}{\sqrt{2\tr(\R^2)}}\\
=&\frac{\bm \upsilon_{\mathcal{M}^c}^\top \R_{\mathcal{M}^c} \bm \upsilon_{\mathcal{M}^c}-N}{\sqrt{2\tr(\R^2)}}+\frac{2\bm \upsilon_{\mathcal{M}^c}^\top \R_{\mathcal{M}^c\mathcal{M}} {\bm \upsilon}_{\mathcal{M}}}{\sqrt{2\tr(\R^2)}}+\frac{2\bm \upsilon_{\mathcal{M}^c}^\top \R_{\mathcal{M}^c\mathcal{M}} {\bm \omega}_{\mathcal{M}}}{\sqrt{2\tr(\R^2)}}\\
&+\frac{{\bm \upsilon}_{\mathcal{M}}^\top \R_{\mathcal{M}} {\bm \upsilon}_{\mathcal{M}}}{\sqrt{2\tr(\R^2)}}+\frac{2{\bm \upsilon}_{\mathcal{M}}^\top \R_{\mathcal{M}} {\bm \omega}_{\mathcal{M}}}{\sqrt{2\tr(\R^2)}}+\frac{{\bm \omega}_{\mathcal{M}}^\top \R_{\mathcal{M}} {\bm \omega}_{\mathcal{M}}}{\sqrt{2\tr(\R^2)}}\\
\doteq &T_1+T_2+T_3+T_4+T_5+T_6
\end{align*}
By the condition (A1), we have $\bm \upsilon_{\mathcal{M}}$ is independent of $\bm \upsilon_{\mathcal{M}^c}$. So $M_1$ is independent of $T_1$. By the results of Theorem \ref{tho3}, we have $M_2$ is asymptotically independent of $T_1$. So we only need to show that $T_i=o_p(1)$ for $i=2,\cdots,5$. By the proof of Lemma \ref{le1}, we have $T_2=o_p(1)$ and $T_4=o_p(1)$ if $m=o(N^{1/2})$. And
\begin{align*}
E(T_3^2)=&O\left(\frac{\omega_{\mathcal{M}}^\top \R_{\mathcal{M}\mathcal{M}^c}\R_{\mathcal{M}^c\mathcal{M}}\omega_{\mathcal{M}}}{\tr(\R^2)}\right)=O(N^{-1}T\tilde{\bm \alpha}^\top \tilde{\bm \alpha})=o(1)\\
E(T_5^2)=&O\left(\frac{\omega_{\mathcal{M}}^\top \R^2_{\mathcal{M}}\omega_{\mathcal{M}}}{\tr(\R^2)}\right)=O(N^{-1}T\tilde{\bm \alpha}^\top \tilde{\bm \alpha})=o(1)
\end{align*}
by the alternative hypothesis (\ref{h12}). So $T_3=o_p(1)$ and $T_5=o_p(1)$. Thus, by the Lemma \ref{small_lemma}, we complete the proof. \hfill$\Box$


\end{document}